\documentclass[useAMS,usenatbib]{mn2e}

\usepackage{graphics}
\usepackage{epsfig}
\usepackage{latexsym}
\newcommand{\be}{\begin{equation}}
\newcommand{\ee}{\end{equation}}
\newcommand{\ba}{\begin{eqnarray}}
\newcommand{\ea}{\end{eqnarray}}
\newcommand{\bc}{\begin{center}}
\newcommand{\ec}{\end{center}}
\newcommand{\lsi}{LS~I~+61$^{\circ}$303}

\title[Long-term monitoring of   \lsi\ with  {INTEGRAL} ]{Long-term monitoring of   \lsi\ with  {INTEGRAL}  }

\author[Zhang et al.]{
S. Zhang$^{1}$\thanks{szhang@ihep.ac.cn}, D. F. Torres$^{2,3}$\thanks{dtorres@ieec.uab.es}, J. Li$^{1}$, Y.P. Chen$^{1}$, N. Rea$^3$ \& J.M. Wang$^{1,4}$\\
$^1$Laboratory for Particle Astrophysics, Institute of High
Energy Physics, Beijing 100049, China\\
$^2$Instituci\'o Catalana de Recerca i Estudis Avan\c{c}ats 
(ICREA) Barcelona,
Spain \\
$^3$Institut de Ci\`encies de l'Espai (IEEC-CSIC),
              Campus UAB,  Torre C5, 2a planta,
              08193 Barcelona, Spain \\
$^4$Theoretical Physics Center for Science Facilities (TPCSF), CAS
                            }

\begin{document}

\date{}

\pagerange{\pageref{firstpage}--\pageref{lastpage}} \pubyear{2008}

\maketitle

\label{firstpage}

  \begin{abstract}
   \lsi\ is one of the few high-mass X-ray binaries that have been recently observed at TeV and GeV energies.
 Here we investigate the hard X-ray spectral and timing properties of this source using the IBIS/ISGRI instrument on-board the {\em INTEGRAL} satellite. We  carry out a systematic analysis based on all available {\em INTEGRAL} data since December 28, 2002 up to April 30, 2008. The total exposure time analyzed amounts to 2.1 Ms, hence more than doubling the previous reported sample. The source is best detected in the 18-60 keV band, with a significance level of 12.0$\sigma$. The hard X-ray data are best fit with a simple power law with a photon index of $\sim$ 1.7$\pm$0.2. We detect a  periodical signal at 27 $\pm$ 4 days,
matching the orbital period of 26.496 days previously reported at other wavelengths. The hard X-rays orbital lightcurve is obtained and compared with those derived at other frequencies.    
\end{abstract}


\section{Introduction}

\lsi\ is among the few binaries observed to date to emit high energy $\gamma$-rays. It hosts a BO Ve main-sequence star (Hutchings \& Crampton 1981; Casares et al. 2005), orbited by a compact object of yet unknown nature, with a period of 26.4960$\pm$0.0026 days (Gregory 2002). The system is located at a distance of about 2 kpc. The orbital zero phase is taken at JD 2443366.775 (Gregory \& Taylor 1978), and the phase values of 0.23$\pm$0.02 (Casares et al. 2005), $\sim$0.275 (Aragona et al. 2009), and 0.30$\pm$0.01 (Grundstrom et al. 2007) represent the periastron passage uncertainty.

\lsi\ has been recently detected as a periodic $\gamma$-ray source by the Major Atmospheric Imaging Cerenkov telescope (MAGIC, Albert et al.
2006) and The Very Energetic Radiation Imaging Telescope Array System (VERITAS, Acciari et al. 2009). A distinctive orbital modulation of the VHE $\gamma$-ray emission was seen, which was found to be anti-correlated with that observed at GeV energies by the {\em Fermi} gamma-ray satellite (Abdo et al. 2009). In the last few years, there has been a burst of activity trying to understand the nature of this source, whether it is composed by a pulsar or a black hole system, and which are the mechanisms  that lead to their multiwavelength behavior (e.g., Dubus 2006, Gupta \& Bottcher 2006, Romero et al. 2007, Sierpowska-Bartosik \& Torres 2009). 

The current observations on \lsi\ are rather sporadic, especially at high energies, which prevent from monitoring the phase evolution on time scale of years. The limited observations at soft X-rays conducted by  {\it XMM-Newton} (Neronov \& Chernyakova 2006; Sidoli et al. 2006), {\em Chandra} (Paredes et al. 2007, Rea et al. 2010), {\em ASCA} (Leahy et al. 1997), {\em ROSAT} (Goldoni \& Mereghetti 1995; Taylor et al. 1996) and {\em Einstein} (Bignami et al. 1981) were too short to cover an entire orbit.  An exception to this were the {\em R}XTE/PCA observations.  {\em R}XTE observed \lsi\ in 1996 March 1 -- 30, and found that the 2 -- 10 keV flux peaks at the orbital phase $\sim$ 0.45 -- 0.6. The detection and non-detection by HEXTE  suggest a similar trend in orbital modulation at 15-150 keV  (Harrison et al. 2000).  Smith et al. (2009) adopted about half year of {\em R}XTE/PCA observations covering the time period from  August 28, 2007 to February 2, 2008 (MJD 54340-54498) and reported a possible two-peak orbital lightcurve at 2--10 keV band and several very short flares (which due to the large FoV of the PCA instrument, could still come from other sources in the field). The orbital phase derived at soft X-rays (0.3-10 keV) peaks at the phase 0.65, close to the apastron (Esposito et al. 2007). This is consistent with the contemporary observations at above hundred GeV by MAGIC (Albert et al. 2009,2009b).

At hard X-rays, although the source was monitored by {\em INTEGRAL} (Chernyakova, Neronov \& Walter 2006), data analyzed till now was not enough to explore the appearance of the orbital period in the power spectrum nor to construct a detailed lightcurve. 
Here we investigate the hard X-ray spectral and timing properties of this source using the IBIS/ISGRI instrument on-board the {\em INTEGRAL} and carry out a systematic analysis based on all available {\em INTEGRAL} data since December 28, 2002 up to April 30, 2008. The total exposure time analyzed amounts to 2.1 Ms thus enhancing the previously published report (Chernyakova, Neronov \& Walter 2006) by taking into account more than 3 years of additional data, more than doubling the previous sample.

\section{Observations and data analysis}

{\em INTEGRAL} (Winkler et al. 2003) is a 15 keV - 10 MeV $\gamma$-ray mission. Its main instruments
are the Imager on Board the {\em INTEGRAL} Satellite (IBIS, 15 keV -- 10 MeV; Ubertini et al. 2003) and the SPectrometer onboard {\em INTEGRAL} (SPI, 20 keV - 8 MeV; Vedrenne et al. 2003). 
These instruments are supplemented by the Joint European X-ray Monitor (JEM-X, 3-35 keV) (Lund et al. 2003) and the Optical
Monitor Camera (OMC, V, 500 -- 600 nm) (Mas-Hesse et al. 2003).  At the lower energies (15 keV -- 1 MeV),
the CdTe array ISGRI (Lebrun et al. 2003) of IBIS has a better continuum sensitivity than SPI.
The satellite was launched in October 2002 into an elliptical orbit with a period of 3 days. Due to the coded-mask design of the detectors, the satellite normally operates in dithering mode, which suppresses the systematic effects on spatial and temporal backgrounds.
The {\em INTEGRAL} observations were carried out in the
so-called individual  SCience Windows (SCWs), with a typical time duration of about 2000 seconds each.
In this work, only IBIS/ISGRI public data are  taken into account; the source is too weak to be detected by JEMX and SPI.
The available {\em INTEGRAL} observations, when \lsi\ had  offset angle less than 14
degrees,  comprised about 875
SCWs, adding up to a total exposure time of $\sim$2.1 Ms (covering rev. 25 -- 667, MJD: 52636 -- 54586).
This total exposure 
then enhances the previously published report (Chernyakova, Neronov \& Walter 2006) by taking into account more than 3 years of additional data, i.e., more than double the previous sample.
The data reduction was  performed by using the standard Online Science Analysis (OSA), version 9.0. The results
were obtained by running the pipeline from the flowchart to the image level, and the spectrum was derived using the mosaic images, as are appropriate for spectral analysis of faint sources. 
The spectra were fitted with {\tt XSPEC} v12.3.1  
and the errors on the model parameters were estimated  at 90$\%$ confidence level.

In order to search for a periodic signal in the lightcurve data, we used the Lomb-Scargle periodogram method (Lomb 1976; Scargle 1982) and followed the procedure described in Farrell et al. (2009).
Power spectra were generated for the lightcurve using the {\tt PERIOD} subroutine  (Press \& Rybicki 1989).
We set oversampling and high-frequency factors, which determine the period range and resolution (Press \& Rybicki 1989),  to two and four, respectively, allowing to search for periodic variability in the 1.1 -- 1000 d range. 
The 99\% white noise significance levels were estimated using Monte Carlo simulations (see e.g. Kong, Charles \& Kuulkers 1998). The 99\% red noise significance levels were estimated using the {\tt REDFIT} subroutine, which can provide the red-noise spectrum via fitting a first-order auto-regressive process to the time series  (Schulz \& Mudelsee 2002, Farrell et al. 2009).\footnote{See ftp://ftp.ncdc.noaa.gov/pub/data/paleo/softlib/redfit}

 \begin{table}
\begin{center}
\label{tab1}
\caption{The flux and significance of \lsi\ obtained from the mosaic images of the combined ISGRI/IBIS observations.  }
\vspace{5pt}
\small
\begin{tabular}{lcccl}
\hline \hline
Energy & \multicolumn{2}{c} { flux } & sig.\\
 (keV)&  ct/s  &  mCrab & $\sigma$  \\
\hline
20-40 &  0.21$\pm$0.02&1.60$\pm$0.16 &10.0\\
\hline
40-100 & 0.16$\pm$0.02&2.10$\pm$0.28 &7.6 \\
\hline
100-300 & 0.07$\pm$0.02&5.00$\pm$1.25 &4.0\\
\hline
18-60 & 0.32$\pm$0.03 &1.60$\pm$0.13 & 12.0 \\
\hline
\hline
\end{tabular}
\end{center}
\end{table}

 \begin{table*}
\begin{center}
\label{tab2}
\caption{Orbitally separated spectra from ISGRI/IBIS observations (the results of our analysis are marked with a star) and a comparison with the previous work by Chernyakova et al. (2006); quoted in the three last columns.  }
\vspace{5pt}
\small
\begin{tabular}{ cccc      ccc}
\hline \hline
   orbital     &  $\Gamma^\star$ (1 $\sigma$)    &          flux$^\star$  (20--60 keV)                                &  effective              &  $\Gamma$ (1 $\sigma$)    &          flux  (20--60 keV)  &  effective                             \\
   phase     &                                                &          (10$^{-11}$ erg cm$^{-2}$ s$^{-1}$) &     exposure$^\star$ (ks)  &                                                &          (10$^{-11}$ erg cm$^{-2}$ s$^{-1}$) &     exposure (ks)  \\
    \hline
 0.4-0.6  & 1.9 $\pm$ 0.2    &    2.84$\pm$ 0.54             &       127    &     1.7$\pm$0.4 & 3.8$\pm$0.6 & 50     \\   
 0.6-0.8  &  1.5 $\pm$0.3    &    2.09$\pm$ 0.52               &        144    &     3.6$^{+1.6}_{-1.1}$  & 3.0$\pm$1.0 & 23      \\ 
 0.8-0.4   & 1.4$^{+0.4}_{-0.3}$ &      1.07$\pm$ 0.36                &       307    &    1.4$\pm$0.3 & 2.4$\pm$0.3 &200                \\  
 whole orbit &     1.7 $\pm$ 0.2 &       1.74$\pm$ 0.26     &        578   &  1.6$\pm$0.2 &  2.5$\pm$0.3 & 273   \\ 
\hline
\hline
\end{tabular}
\end{center}
\end{table*}

The most significant {\it INTEGRAL} detection of \lsi\ is derived from combining all the ISGRI data, and is found at a significance level of $\sim$ 12.0$\sigma$ in the 18--60 keV band (see Figure 1). This significance is consistently higher than that reported by Chernyakova, Neronov \& Walter (2006, of 8.1$\sigma$) at the 22--63 keV, based on 600 SCWs (covering rev. 25-288, from 2003 January to 2005 March, and producing a corrected exposure of 273 ks; as opposed to ours of about 578 ks). 
Table 1 shows the details on the imaging results in 4 energy bands: 20-40, 40-100, 100-300, and 18-60 keV. The  neighboring source QSO 0241+622 is detected at 3.4 mCrab level in 18-60 keV, implying a $\sim 25\sigma$ detection.  This source is only about 1.4 degrees away and it is usually a problem for the analysis of \lsi\ with non-imaging detectors. The energy spectrum of the source is produced  combining all the ISGRI data in the 13 keV -- 1.0 MeV band. Since we note that the
sky region around LS I +61$^o$ 303 is rather crowded and \lsi\ itself is relatively faint in hard X-rays, it is necessary to obtain the energy spectrum from the mosaic images (see also Chernyakova \& Neronov 2008). The energy spectrum is well fitted by a power law shape, with a photon index of 1.7$\pm$0.2 and normalization of 3.1$^{+2.7}_{-3.1}$$\times$10$^{-3}$ph cm$^{-2}$ s$^{-1}$ keV$^{-1}$. The flux we derive in the 20-200 keV band is 2.8 $\times$10$^{-11}$ergs cm$^{-2}$ s$^{-1}$. The reduced $\chi^2$ for the fit is 0.4 under 7 dof. 
The orbitally separated results of the spectral fitting can be seen in Table 2. 
These results show that our spectral fittings are mostly consistent with those published previously by Chernyakova, Neronov \& Walter (2006). They also present a large improvement, especially for the phase interval 0.6-0.8, where the error bar in the photon index is largely reduced from $>1$ to $0.3$.

\begin{figure}
\centering
 \includegraphics[angle=0, scale=0.4]{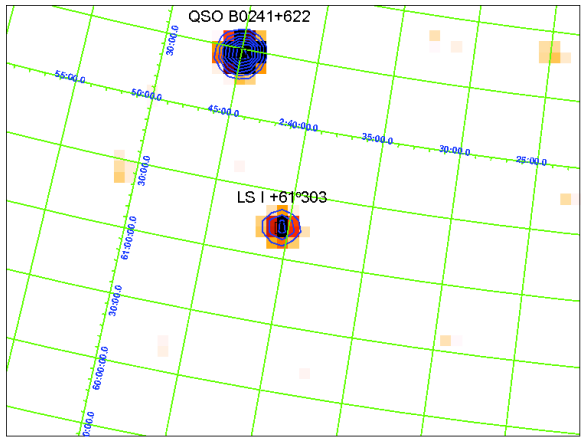}
      \caption{The mosaic image of the \lsi\ sky region derived by combining all ISGRI data at the 18 -- 60 keV band. The strongest source is QSO 0241+622 (at the upper part of the image) and the relatively faint one is \lsi. The significance level is given in the color scale. The contours start at a detection significance level of 5 $\sigma$, with a step of 3 $\sigma$. }
         \label{lc_flux}
\end{figure}

\begin{figure}
\centering
 \includegraphics[angle=270, scale=0.35]{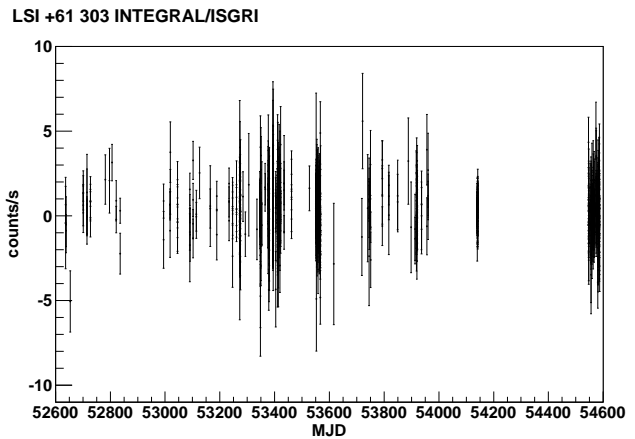}
      \caption{The ISGRI lightcurve at 18-60 keV. The bin size is one SCW and the error bar is 1$\sigma$.  }
         \label{lc_flux2}
\end{figure}


The ISGRI lightcurve of the source is obtained in the 18--60 keV range with a time bin following the SCWs (see Figure 2). A  fit to this lightcurve with a constant produces a value of $\chi^2\sim$ 1.2 (1058/874), indicating a deviation at a significance level of 4$\sigma$. We then investigated whether there is a recognizable orbital variability in this data using timing analysis. The orbital lightcurve is then made by using the radio period and a reference time at MJD=43366.275 (Gregory 2002) at the hard X-rays, so that the absolute orbital phases can be directly compared  with those at other frequencies (and we do this below). We find that at the hard X-rays   the maxima is around phase 0.5-0.6 whereas the periastron of the system is at around 0.25-0.3 (Casares et al. 2005, Grundstrom et al. 2007, Aragona et al. 2009). 
The Lomb-Scargle power spectrum based on the ISGRI lightcurve  is presented in Figure 3.
The maximum of this power spectrum is located at around 27 days with a power density of 12.8 and an error of 4 days based on Monte Carlo simulations. This periodical signal matches the orbital period of   26.4960 days, as reported  in radio (Gregory 2002).
To obtain the error in the period determination we have
sampled the light curve that was used to produce the power spectrum. The flux and flux error of each bin of this light curve are regarded as the mean and the deviation of a Gaussian. Accordingly, for each bin, a Monte Carlo flux is obtained to form a new light curve. In the power density of each so-generated lightcurves, the peak around 26.5 days is taken. The width of this peak distribution provides one sigma error of the measured orbital period 26.72 days from the real ISGRI ldata. Over-plotted in Figure 3
are the white noise and the red noise both at 99\% confidence level.  The comparison of our {\em  INTEGRAL} lightcurve with those obtained at other frequencies is shown in Figure 4, which is further discussed next.

\begin{figure}
\centering
  \includegraphics[angle=270, scale=0.33]{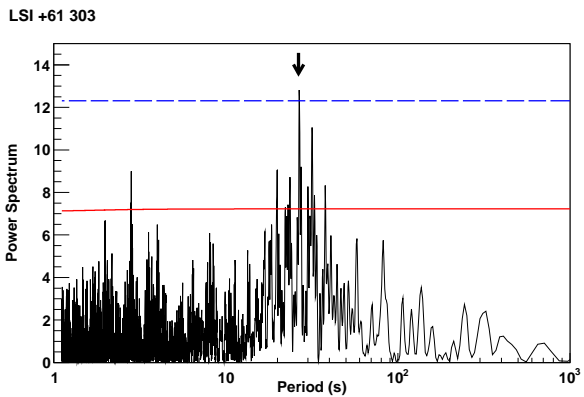}
      \caption{The Lomb-Scargle power spectrum  based on the ISGRI lightcurve (18 -- 60 keV).
      The white noise (dashed line) and the red noise (solid line) at the 99\% confidence level are plotted. The \lsi\ orbital period of 26.496 days is indicated with an arrow.}
         \label{lc_HID0}
\end{figure}

\begin{figure*}
\centering
 \includegraphics[angle=0, scale=0.45]{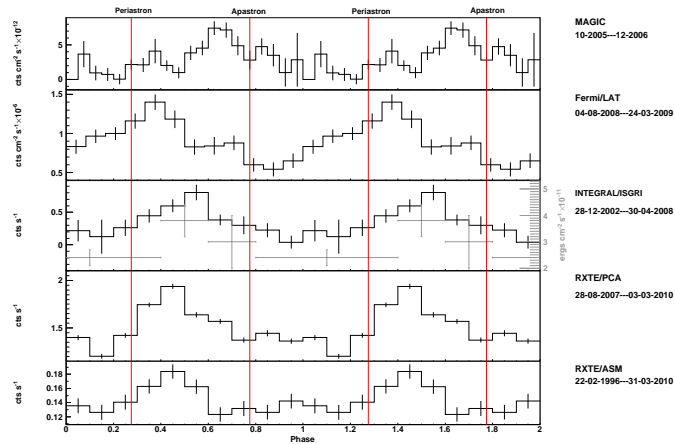}
      \caption{The orbital lightcurves in a multi-frequency context, from X-rays  to GeV (Abdo et al. 2009) and TeV gamma-rays (Albert et al. 2009). All the plots in this paper have adopted the periastron passage as 0.275 and apastron passage 0.775 as shown by the vertical solid  lines (Aragona et al. 2009).
         }
         \label{lc_HID_all}
\end{figure*}

\section{Discussion}

We have carried out a systematic analysis based on all available {\em INTEGRAL} data since December 28, 2002 up to April 30, 2008. The total amount of time analyzed was $\sim$2.1 Ms, thus enhancing the previously published report (Chernyakova, Neronov \& Walter 2006) by taking into account more than 3 years of additional data, more than doubling the previous sample. Corrected exposures   after cuts changed from 273 to 578 ks.  The difference in the amount of data analyzed allows for a much more detailed {\em INTEGRAL} lightcurve to be constructed, as can be  visually seen in panel 3 of Figure 4 above, where we show the Chernyakova, Neronov \& Walter (2006) 20-60 keV lightcurve in the background, and our current results superposed.
Our analysis of  {\em INTEGRAL}/ISGRI data results in the first direct orbital period determination through a power spectrum analysis of the hard X-ray data of \lsi, although not yet truly significant, and in a  spectrum at hard X-rays which is similar to the previous report (Chernyakova, Neronov \& Walter 2006), but with detection significance improved to 12.0$\sigma$, meaning that the source did not show significant changes in hard X-rays along the last five years. 

Our analysis also confirms that in the case of \lsi\ there is no high-energy cut-off found at energies below 100 keV, as it would be common in conventional accretion scenarios. 
Indeed, if the system is an accreting neutron star or black hole, one expects to find a cut-off power-law spectrum in the hard X-ray band with a cut-off energy normally at 10 -- 60 keV for neutron stars (e.g., Filippova et al. 2005) and at  $\sim$100 keV for black holes (McClintock \& Remillard 2003).  Even when one can think of situations where the accretion disk is masked by a jet component, producing a featureless power law, if the accretion flow and the jet would give comparable contributions in the hard X-ray band, one would still expect at least a feature in the INTEGRAL energy range, due to the disappearance of of contribution of the disk at high energies. Such is not observed, and thus, in case an accretion disk is present, its contribution should be sub-dominant with respect to the jet one. The flux we derive in the  20-200 keV band is 2.8 $\times$10$^{-11}$ergs cm$^{-2}$ s$^{-1}$.  
If most of it comes from a jet, and since in the scenario being discussion the whole source activity would be powered by accretion, the luminosity ($L_{jet} \sim  10^{35}$ erg s$^{-1}$, of the order of the gamma-ray luminosity) could only be explained by a large beaming factor. From $\sim$50 mas resolution radio images of \lsi\ obtained with Multi-Element Radio Linked Interferometer (MERLIN), extended, apparently precessing, radio emitting structures at angular extensions of $0.01-0.05$ arcsec have been reported by Massi et al. (2001, 2004). However, recent Very Long Baseline Array (VLBA)  imaging obtained by Dhawan et al. (2006) over a full orbit of \lsi\ has shown the radio emission to come from angular scales smaller than about 7~mas (which is a projected size of 14 AU at an assumed distance of 2~kpc). No large features or high-velocity flows were noted in any of the observing days, which implies at least a non-permanent jet nature. Changes within 3 hours were found to be insignificant, so the velocity of the outflow can not be much over 0.05$c$.
The MAGIC collaboration also conducted a radio campaign (in concurrence with TeV observations) to test these results using again   MERLIN in the UK, the European VLBI Network (EVN), and the VLBA in the USA (Albert et al. 2008). The results obtained by radio imaging at different angular scales show that the size of the radio emitting region of \lsi\ is constrained to be below $\sim$6~mas ($\sim$12 projected AU), and the presence of persistent jets above this scale is therefore excluded. As in the case of Dhawan et al. (2006), these observations have shown a radio-emitting region 
extending east-southeast from the brighter, unresolved emitting core. The outflow velocity implied by these observations is 
$\sim 0.1c$. These values of $\beta$ would result too low to significantly increase the luminosity ($L_{jet} \propto \delta^4$) even if strong jet precession is assumed. Finally, search for lines with the Chandra observatory yielded to no evidence for absorption or emission features in the source averaged and time-resolved spectra (Rea et al. 2010). Putting the {\em INTEGRAL} observations in this context, we find no obvious support for the accretion model in our data.

As can be seen in Figure 4, the hard X-ray emission peaks in phase around 0.5, and 
in order to compare with the orbital phase derived at $\gamma$-rays, we have also generated the orbital light curves of {\em R}XTE/PCA and {\em R}XTE/ASM using the currently available observations.  The orbital peaks of the soft and hard X-rays are close in phase
(0.4-0.6) but not co-aligned. Thus, with low statistics, the hard X-rays may appear following the overall orbital modulation trend of the soft X-rays  (e.g. see Harrison et al. 2000). We also show in Figure 4 the lightcurves obtained by Fermi (Abdo et al. 2009, 8 months of observations since launch in June 2008), and MAGIC (several orbits put together, along few years of observations, as described in Albert et al. 2009). Direct inspection of these lightcurves seem to indicate that the hard X-ray component peaks in between the average soft X-ray and TeV lightcurves, but significant variability (e.g., orbit to orbit variability is already seen in MAGIC and VERITAS data, Albert et al. 2009, Acciari et al. 2009) may preclude to obtain conclusions in non-simultaneous observations.

\section*{Acknowledgments}

This work was subsidized by the National Natural Science Foundation of China, the CAS key Project KJCX2-YW-T03, and 973 program 2009CB824800. J.-M. W. thanks the Natural Science Foundation of China for support via NSFC-10325313, 10521001 and 10733010. DFT and NR acknowledges support from grants AYA2009-07391 and SGR2009-811.

\label{lastpage}

\begin{thebibliography}{}

\bibitem[Abdo  (2009)]{Abdo09} Abdo A. et al.,  2009, ApJ, 701, 123
\bibitem[Acciari  (2009)]{Acciari09} Acciari V.A. et al.,  2009, ApJ, 700, 1034
\bibitem[Albert (2006)]{Albert06} Albert J. et al., 2006, Science, 312, 1771 
\bibitem[Albert (2008)]{Albert08} Albert J. et al., 2008, ApJ, 684, 1351 
\bibitem[Albert  et al. (2009)]{Albert09}  Albert  J. et al.,  2009, ApJ, 693, 303
\bibitem[Albert (2009b)]{Albert09b} Albert J. et al., 2009b, ApJ Letters, 706, 27  
\bibitem[Aragona  et al. (2009)]{Aragona09} Aragona C. et al.,  2009, ApJ, 698, 514

\bibitem[Bignami et al.   (1981)]{Bignami81}  Bignami G.F.B. et al., 1981, ApJ, 247, L85


\bibitem[Casares et al. (2005)]{Casares05} Casares J. et al., 2005, MNRAS, 360, 1105  
 

\bibitem[Chernyakova et al. (2005)]{Chernyakova05} Chernyakova M., Neronov A., Walter R., 2006, MNRAS, 372, 1585
 \bibitem[Chernyakova et al. (2008)]{Chernyakova08} Chernyakova M.,  Neronov A., 2008, IBIS Analysis User Manual


\bibitem[Dhawan (2006)]{Dhawan06}  Dhawan V., Mioduszewski A.,  Rupen M. 2006, VI Microquasar Workshop, PoS 52.1

\bibitem[Dubus (2006)]{Dubus06} Dubus G., 2006, A\&A, 456, 801   
\bibitem[Esposito  et al. (2007)]{Esposito07} Esposito P. et al., 2007, A\&A, 474, 575
\bibitem[Farrell  et al. (2009)]{Farrell09} Farrell S.A., Barret D., Skinner G.K., 2009, MNRAS, 393, 139    

\bibitem[Filippova (2005)]{Filippova05} Filippova E. V., Tsygankov S. S., Lutovinov A. A.,  Sunyaev R.
A., 2005, Astr. Lett., 31, 729.
\bibitem[Goldoni et al. (1995)]{Goldoni95} Goldoni P.,  Mereghetti S., 1995, A\&A, 299, 751  
 
 
\bibitem[Gregory \& Taylor (1978)]{Gregory78} Gregory P.C.,  Taylor A.R., 1978, Nature, 272, 704
\bibitem[Gregory (2002)]{Gregory02} Gregory P.C., 2002, ApJ, 525, 427\ 
\bibitem[Grundstrom et al. (2007)]{Grundstrom07} Grundstrom E. D. et al., 2007, ApJ, 656, 437
\bibitem[Gupta (2006)]{Gupta06} Gupta S.,  Bottcher M., 2006, ApJ, 650, L123
 

\bibitem[Harrison et al. (2000)]{Harrison00} Harrison F.A. et al., 2000, ApJ, 528, 454 
\bibitem[Hutchings \& Crampton  (1981)]{Hutchings81}  Hutchings  J.D.,  Crampton D., 1981, PASP, 93, 486
 
 
\bibitem[Kniffen et al. (1997)]{Kniffen97} Kniffen D.A. et al., 1997, ApJ, 486, 126  


\bibitem[Kong  et al. (1998)]{Kong98}  Kong A. K. H., Charles P. A.,  Kuulkers E.,  1998, New Astron., 3, 301





\bibitem[Leahy et al.    (1997)]{Leahy97} Leahy D.A. et al., 1997, ApJ, 475, 823  



\bibitem[Lebrun et al. (2003)]{lebrun03} Lebrun F. et al., 2003, A\&A, 411, L141
\bibitem[Lomb  (1976)]{Lomb76}  Lomb N. R., 1976, Ap\&SS, 39, 447
\bibitem[Lund et al. (2003)]{lund03} Lund N. et al., 2003, A\&A, 411, L231


\bibitem[Massi (2001)]{Massi01}  Massi M.,  Rib\'o M., Paredes J. M., Peracaula M.,  Estalella M., 2001, A\&A, 376, 217
\bibitem[Massi (2004)]{Massi04}  Massi M.  et al., 2004, A\&A, 414, L1

\bibitem[Mas-Hesse et al. (2003)]{mas-hesse03} Mas-Hesse J. M. et al., 2003, A\&A, 411, L261
\bibitem[McClintock (2003)]{McClintock03} McClintock J.E., Remillard R.A., 2003, in Compact Stellar X-ray Sources, eds. W.H.G. Lewin and M. van der Klis, Cambridge
University Press 


\bibitem[Paredes   et al. (2007)]{Paredes07}  Paredes J.M. et al., 2007, ApJ, 664, L39
\bibitem[Press   et al. (1989)]{Press89}  Press W. H., Rybicki G. B.,  1989, ApJ, 338, 277



\bibitem[Rea (2010)]{Rea10} Rea N. et al., 2010, MNRAS, 716 in press 
(arXiv:1002.2223) 

\bibitem[Romero (2007)]{Romero07} Romero G., Okazaki A. T., Orellana M.,  Owocki, S. P., 2007, A\&A, 473, 15

\bibitem[Scargle   et al. (1982)]{Scargle82}  Scargle J. D.,  1982, ApJ, 263, 835
\bibitem[Schulz  et al. (2002)]{Schulz02}  Schulz M.,  Mudelsee M.,  2002, Comput. Geosci., 28, 421
\bibitem[Sidoli  et al. (2006)]{Sidoli06} Sidoli L. et al., 2009, A\&A, 459, 901   
\bibitem[Sierpowska-Bartosik (2009)]{Sierpowska-Bartosik09} Sierpowska-Bartosik A.,  Torres D. F., 2009, ApJ, 693, 1462
\bibitem[Smith   et al. (2009)]{Smith09} Smith A. et al., 2009, ApJ, 693, 1621  

\bibitem[Taylor et al. (1996)]{Taylor96} Taylor A.R. et al., 1996, A\&A, 305, 817  

\bibitem[Ubertini et al. (2003)]{ubertini03} Ubertini P. et al., 2003, A\&A, 411, L131

\bibitem[Vedrenne et al. (2003)]{vedrenne03} Vedrenne G. et al., 2003, A\&A, 411, L63

\bibitem[Winkler et al. (2003)]{winkler03} Winkler C. et al., 2003, A\&A, 411, L1
 


\end{thebibliography}
\end{document}